# Mesh Router Nodes placement in Rural Wireless Mesh Networks

Jean Louis Fendji Kedieng Ebongue*, Christopher Thron**, Jean Michel Nlong*, Karl-Heinz Rodiger***

*The University of Ngaoundéré
CAMEROON

lfendji@univ-ndere.cm, jmnlong@univ-ndere.cm

** Texas A&M University Central Texas

USA

thron@ct.tamus.edu

*** Universität Bremen

GERMANY

roediger@informatik.uni-bremen.de

**RÉSUMÉ.** Le problème de placement de nœud routeur dans les réseaux maillés sans fil est connu comme étant NP difficile. Dans ce papier, le problème est adressé sous une contrainte de modèle de réseau adapté aux zones rurales où nous observons généralement une population clairsemée à faible densité. Nous considérons premièrement la zone initiale comme étant décomposée en zones élémentaires qui peuvent être optionnelles en couverture ou non, et où on peut placer un nœud ou non. Nous proposons par la suite un algorithme basé sur l'approche du Metropolis pour assurer la couverture. L'évaluation de l'algorithme proposé sur une instance de réseau a donné un pourcentage de couverture proche de 100 avec un nombre de routeur optimal.

**ABSTRACT.** The problem of placement of mesh router nodes in Wireless Mesh Networks is known to be a NP hard problem. In this paper, the problem is addressed under a constraint of network model tied to rural regions where we usually observe low density and sparse population. We consider the area to cover as decomposed into a set of elementary areas which can be required or optional in terms of coverage and where a node can be placed or not. We propose an effective algorithm to ensure the coverage. This algorithm is based on metropolis approach. We evaluated the proposed algorithm on an instance network. A close to 100 percent coverage with an optimal number of routers showed the efficiency of our approach for the mesh router node placement problem.

# 1. Introduction

Wireless Mesh Networks (WMNs) [1] are composed of mesh nodes connected in a mesh topology. This kind of network based on WiFi technology is an appealing solution to bridge the digital divide observed between rural and urban regions. Especially in developing countries, WMNs can play a crucial role in the national development; since rural activities like farming and agriculture still remain the pillars of the economy in these countries. The success of this kind of network is due to the low cost of the Wifi technology when comparing to other (especially WiMax) and the continuous capacity improvement of this technology in terms of throughput and coverage. WMN in rural region is usually composed of one gateway which connects the network to Internet, and a set of mesh routers (MRs) and mesh clients. Similar to normal routers, MRs incorporate some functionality to support mesh networking.

The performance in terms of connectivity and coverage of a WMN relies on an optimal placement of MRs. In rural regions, especially in developing countries, a real concern when designing such a network is the overall cost. By its nature, the problem of mesh node placement requires a multi objective approach; since it is a NP-hard combinatorial optimization problem which cannot be solved in polynomial time. Usually these objectives seem to be contradictory like: minimising the number of MR while keeping or extending the coverage.

In this paper, we address the problem under a constrained network model tied to rural regions where we usually observe low density and sparse population. We first decompose the area to cover into elementary areas which can be required (school, hospital…) or optional (farm, road…) in terms of coverage and where a node can be placed or not. The objectives here are: (1) to minimise the number or MR and the coverage of optional areas; (2) to maximise the coverage of required areas. We firstly define the network model and provide a formulation of the placement problem in rural region. Afterwards, we propose an effective heuristic to obtain a close to optimal coverage of required areas using a minimal number of router. The algorithm is based on metropolis approach. Finally, we evaluate the proposed algorithm on an instance network using Scilab 5.4.0. A close to 100 percent coverage with a minimal number of routers shows the efficiency of our approach for the mesh router node placement problem.

The rest of the paper is organised as follows: In section 2, we briefly present the previous work in the literature. In section 3, we give the network model and a formulation for the placement problem. The simulated annealing approach for this problem is described in section 4. In section 5, we present the experimental setup to evaluate our approach and discuss the results. We finally conclude the paper in section 6.

## 2. Related Work

The most of work in WMN planning done in rural regions could be considered as partial design since they depend on existing gateway(s).

The mesh router node placement is a crucial aspect of the network design and it depends on the topology of the region. According to the network model and the problem statement, different approaches have been proposed to solve the problem of node placement in WMN. Since this problem is known to be hard [2], search techniques and meta-heuristic are usually used [3, 4, 5, 6, 7]. The region to be covered, usually called the universe, can be considered as continuous (a whole region), discrete (a set of predefined positions) or network (undirected weighted graph).

In [3], the mesh nodes placement problem is tackled using annealing approach. It considers the version of the mesh node placement problem where: given a 2D area where to distribute a number of MR nodes and a number of mesh client nodes of fixed positions (of an arbitrary distribution) they have to find a location assignment for the MRs that maximizes the network connectivity and client coverage.

In [4], the authors study efficient MR placement in WMN. Their MR placement problem is the determination of a minimum set of positions among the candidate positions in such a way that the MRs situated in these positions cover the given region.

Previous works in mesh node placement usually addressed urban region models with a dense population and a whole region to cover.

## 3. Network Model and Problem Statement

### 3.1. Network Model

In rural region, there is no need to cover a whole region. A given region is usually composed of sparse areas of interest (IA) where the signal must be spread (like a market, a school, a hospital…); optional area (OA) where the signal can be spread or not (great farms for example) and where we can place a node and finally prohibiting area (PA) where a node cannot be placed (a lake or a road). The network usually contains only one gateway (IGW) generally fixed and connected to Internet by Satellite. We consider routers with omni-directional antenna and assume them to have the same coverage so that a router can be represented by a circle.

To be more realistic, the area to cover is modelled as a two-dimensional irregular form in a two-dimension coordinate plane. We consider the smallest rectangle that can contain the irregular form. Therefore, we assume that this rectangle is decomposed in small square forms called elementary area (EA) in other to obtain a grid. Hence, we obtain a set of elementary area of interest (IEA) and a set of prohibitive elementary area (PEA). Thinking like this, we can define different two-dimensional matrices to characterise each EA. Let consider the matrices: Cover defining whether an EA requires coverage or not; and Place whether in an EA we can place a node. Therefore, an EA at position (x, y) can be characterised by (1), (2) and (3):

$$Cover(x,y) = \begin{cases} 0 \rightarrow coverage\ not\ required \\ 1 \rightarrow coverage\ required \end{cases} \quad (1)$$

$$Place(x,y) = \begin{cases} 0 \rightarrow cannot\ place\ a\ node \\ 1 \rightarrow can\ place\ a\ node \end{cases} \quad (2)$$

$$CoverDepth(x,y) = \begin{cases} 0 \rightarrow no\ coverage \\ x \rightarrow covered\ by\ x\ routers \end{cases} \quad (3)$$

Figure 1 illustrates the result of a decomposition of a region into a set of EA.

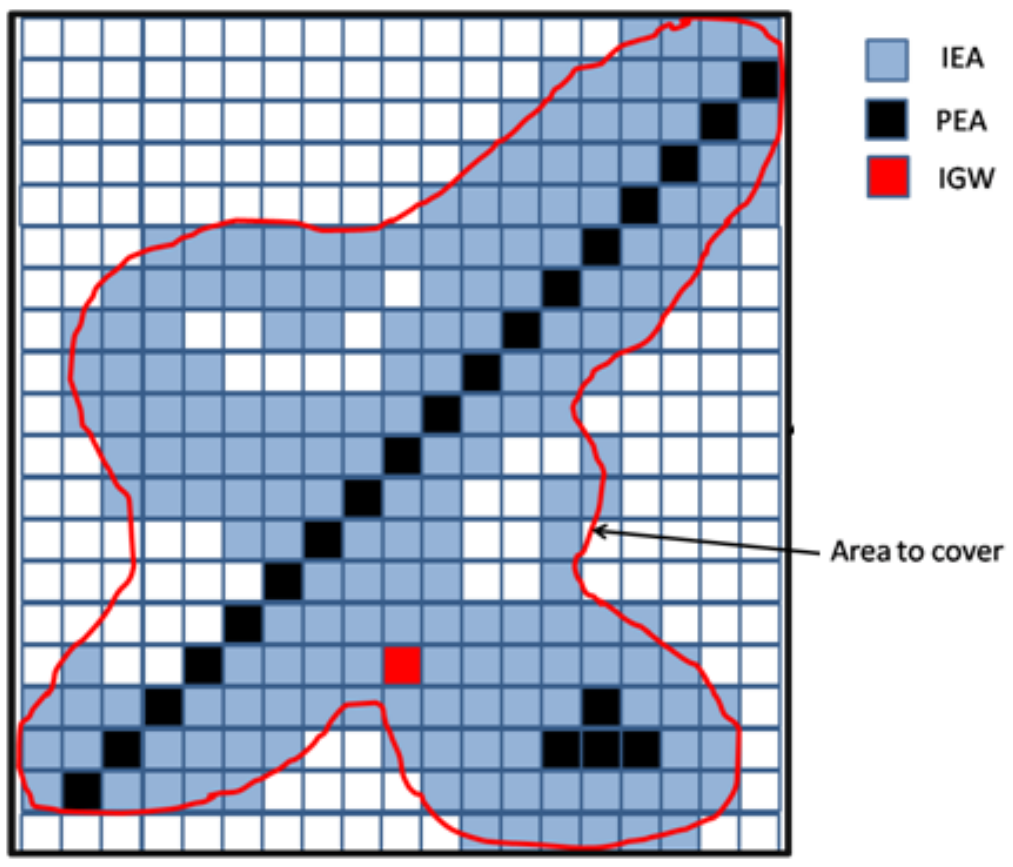

Figure1: An example of region decomposed in EA

The population is not so dense like in urban region; thus, we consider a uniform repartition of clients, which means each EAI has the same number of client. We consider routers to have the same radius (r). This radius is expressed in number of EA. r= 4 means that the radius stretches over 4 EAs.

Let p an EA at position (x, y). If a MR is located in p, then the set of EA cover by this MR is given by (4).

$$\forall(x,y), x^2 + y^2 < r^2 \quad (4)$$

### 3.2. Problem Statement

The main concern when deploying WMN in a rural region is the overall cost. This cost is influenced by the number of MR. The more the region to cover is big, the more we need router and the cost is increased. So to minimise cost, we need to cover only areas of interest. Therefore, the MR placement problem in rural regions can be described as the determination of minimum set of positions which maximises the coverage of required areas minimises the coverage of optional areas while minimising the number of MRs.

## 4. Metropolis approach

### 4.1. Algorithm

Metropolis algorithm is a meta-heuristic designed to solve global optimization problems by finding a good approximation to the global optimum. Metropolis algorithm is a specialisation of simulated annealing algorithm with a non-variant temperature. A pseudo code for metropolis algorithm is the following:

```
Set T
S := Initial Solution()
V := Evaluate(S)
while (stopping condition not met) do
        St := Generate(S)
        Vt := Evaluate(St)
        if Accept(V,St,T ) then
                S:= St
                V := Vt
        end if
end while
return S
```

T is the temperature, S the solution, St the temporary solution, V the value of the fitness function and Vt the temporary value of V.

### 4.2. Particularisation of the algorithm

#### 4.2.1. Algorithm parameters

**Initialisation:** The first step is to determine the number of router for a given region. The minimum number of router is given by (5).

Because this minimal number cannot ensure the coverage and the connectivity of the required areas (since routers should overlap), we use an initial number of routers given by (6).

$$\boldsymbol{nr_{min} = \lceil \sum Cover\ (x, y)\ \ /(r^2 * 3.14) \rceil} \quad \textbf{(5)}$$

$$\boldsymbol{nr_{min} < nr_{init} < 2 * nr_{min}} \qquad \textbf{(6)}$$

During this phase, routers are place randomly in the region only on areas of interest. For each router we randomly select an EA. we check if Cover(EA)=1 and Place(EA)=1 then the current router can be placed there. Otherwise, we continue by selecting and EA. The initialisation ends when all routers are placed with Cover(EAi)=1 and Place(EAi)=1.

**Movement:** We define a set of movement and we move only one router at the same time. The movement is randomly selected. A movement from EAi to EAj is accepted if the Cover(EAj)=1.

**Fitness function:** The evaluation of fitness function consists to count the number of covered IEA. This is done by (7) after the initialisation. To be more efficient we calculate only the change in the coverage. Since we move only one router at the same time, we consider the EAs of this router which are concerned by the movement.

$$f = \sum sign(CoverDepth.* \ Cover) \quad (7)$$

$$f_{i+1} = f_i + \Delta f_{i \to i+1} \qquad (8)$$

**Acceptability criteria:** The main difference between Metropolis and Hill Climbing is that even if $\Delta f_{i \to i+1}$ is negative, the movement is accepted with a certain probability influenced by the temperature T (9) with x a random number such as $0 < x < 1$.

$$rand(x) < \ exp(T * \Delta f_{i \to i+1}) \quad (9)$$

**Stopping condition:** If the value of the fitness function does not improve after a certain number of iteration (nbtostop), we supposed having reached the optimal.

### 4.2.2. The optimal number of router

After ensuring a desire percentage of coverage, the next objective is to minimise the number of MR while keeping this percentage. We will remove one router each time and perform movements with the rest. If the desired coverage percentage is satisfied, we continue to remove until it goes down the threshold. Therefore, we consider the previous number and placement of router to be optimal. To remove a router, three strategies can be used: (1) Remove circle with minimum single-coverage; (2) Remove circle with minimum coverage and (3) Remove circle with maximum over-coverage.

## 5. Experimental results

To evaluate our proposed algorithm, we consider a grid of 200x200 with r=8. The unity is the size of an EA. If size(EA)=20m, the grid will 4Km x 4km=16km² and the radius r=160m which is realistic. The other parameters are T=0.1, nbtostop=500, $nr_{Init}=1.4*nr_{min}$. We randomly generate a region with areas of interest and prohibitive areas. Figure 2 shows the initial area to cover. White cells represent areas of interest. Figure 3 illustrates a placement for the optimal number of MR. Blue cells are covered by one router, red cells are covered by two routers and white cells by three.

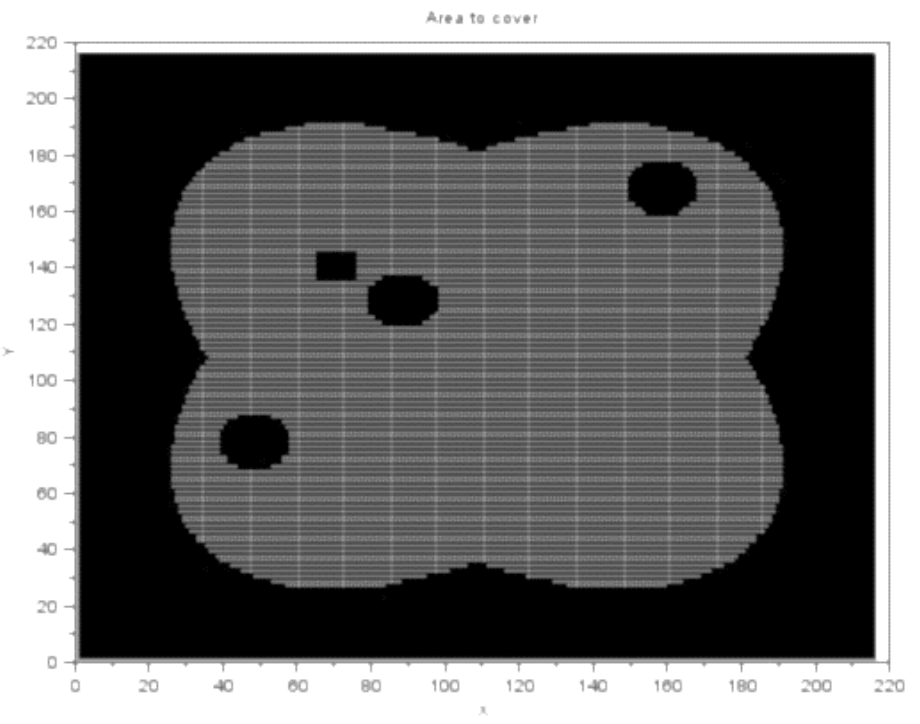

Figure 2: Random initial area to cover

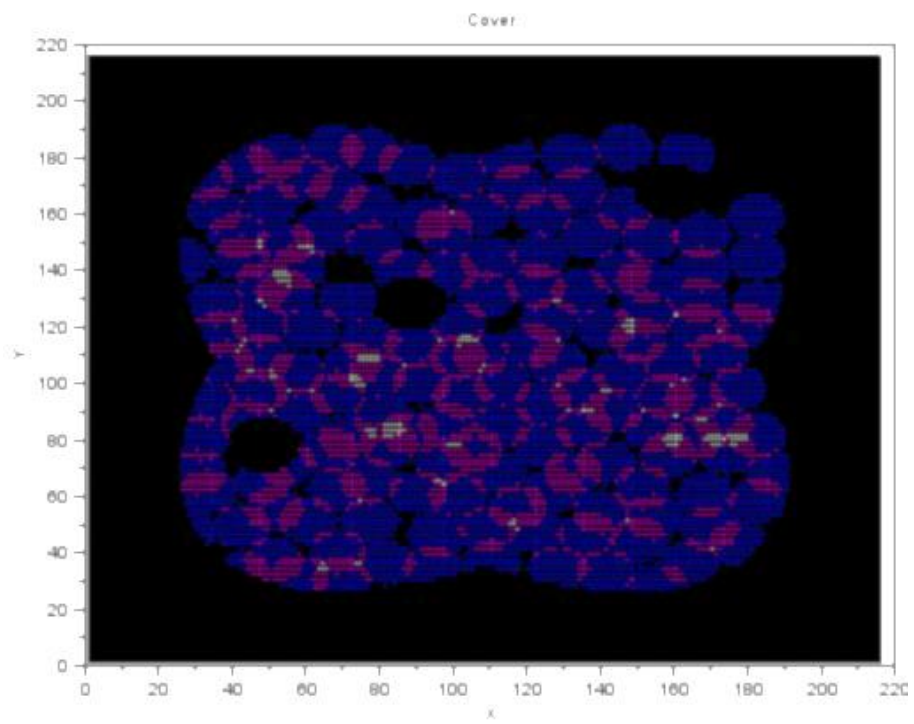

Figure 3: Placement with $nr_{opt}$ MRs

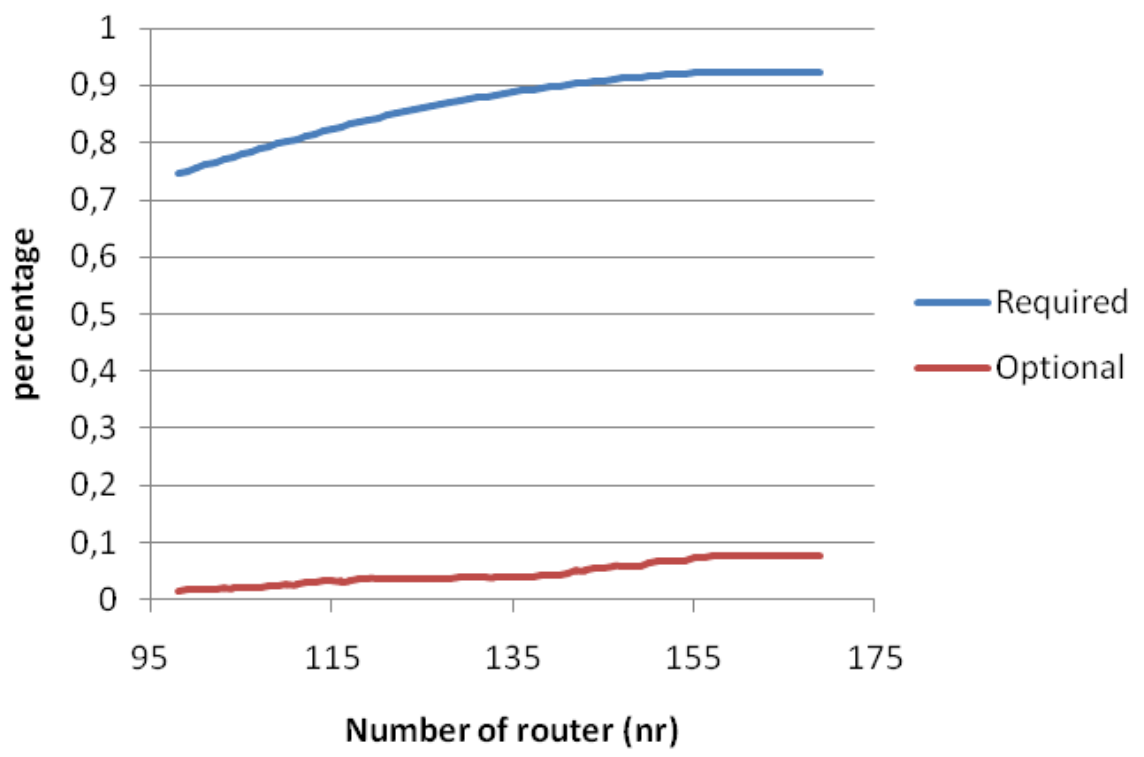

Figure 4: Percentage of coverage

We observe from Figure 4 that the maximal number is $nr_{max}=1.33*nr_{min}$, with a percentage of required coverage between 0.92 and 0.95 and optional coverage <0.08. The percentage cannot longer increase beyond this maximal number of routers. We also observe an optimal number of mesh router $nr_{opt}=1.25*nr_{min}$, while considering $\Delta percentage\left(nr_{max} \rightarrow nr_{opt}\right) < 0.01$. In this case, optional coverage is <0.07. With the $nr_{min}$, we obtain a percentage of required coverage $\cong 0.84$ and optional coverage <0.04.

## 6. Conclusion and future work

This paper has presented a metropolis approach for mesh router node placement in rural WMN. Experimental results showed the efficiency of our approach to solve the problem of MR placement in rural areas while determining an optimal number of MRs. In fact, we obtained a required coverage between 0.92 and 0.95 and an optional coverage less than 0.08. The optimal number of MR is $nr_{opt}=1.25*nr_{min}$.

In this work, we did not consider cases where an area of interest is disjointed from others; because this kind of situation usually results in a separated mesh network topologies. So, besides improving the algorithm in order to obtain a percentage very close to 100, we will investigate on the case of disjointed areas of interest.

## 7. Bibliography

[1] I. F. Akyildiz, X. Wang, and W. Wang. Wireless mesh networks: a survey. Computer Networks 47(4) (2005), 445-487.

[2] E. Amaldi, A. Capone, M. Cesana, I. Filippini, F. Malucelli. Optimization models and methods for planning wireless mesh networks. Computer Networks 52 (2008) 2159-2171.

[3] Xhafa, F., A. Barolli, C. Sánchez, L. Barolli. A simulated annealing algorithm for router nodes placement problem in Wireless Mesh Networks. Simulation Modelling Practice and Theory, In Press, 2010.

[4] J. Wang, B. Xie, K. Cai and D.P. Agrawal. Efficient Mesh Router Placement in Wireless Mesh Networks, MASS 2007, Pisa, Italy (2007).

[5] Xhafa, F., C. Sanchez, and . L. Barolli, Genetic Algorithms for Efficient Placement of Router Nodes in Wireless Mesh Networks in Advanced Information Networking and Applications (AINA), 2010 24th IEEE International Conference on p. 465-472.

[6] Xhafa, F., C. Sanchez, and . L. Barolli, Ad Hoc and Neighborhood Search Methods for Placement of Mesh Routers in Wireless Mesh Networks Distributed Computing Systems Workshops, 2009. ICDCS Workshops '09. 29th IEEE International Conference on p. 400-405.

[7] De Marco, G, MOGAMESH: A multi-objective algorithm for node placement in wireless mesh networks based on genetic algorithms, in Wireless Communication Systems, 2009. ISWCS 2009. 6th International Symposium on p. 388 – 392.